# Vehicle clustering and resource scheduling scheme based on collaborative perception of wireless resources

Jingbin Lu
*School of Computer Science and Engineering, Wuhan Institute of Technology*
Wuhan 430205, China
lujingbin@stu.wit.edu.cn

Zhonghui Pei*
*School of Computer Science and Engineering, Wuhan Institute of Technology*
Wuhan 430205, China
peizhonghui@wit.edu.cn
*Corresponding author

Long Xie
*School of Computer Science and Engineering, Wuhan Institute of Technology*
Wuhan 430205, China
xielong@stu.wit.edu.cn

Zhengxiong Hu
*School of Computer Science and Engineering, Wuhan Institute of Technology*
Wuhan 430205, China
huzhengxiong@stu.wit.edu.cn

Run Wang
*School of Computer Science and Engineering, Wuhan Institute of Technology*
Wuhan 430205, China
wangrun@stu.wit.edu.cn

Yuntao Wu
*School of Computer Science and Engineering, Wuhan Institute of Technology*
Wuhan 430205, China
ytwu@wit.edu.cn

***Abstract*—Cellular Vehicle-to-Everything(C-V2X) sidelink communication allows autonomous resource selection via distributed scheduling without base station coverage. However, the limited perception of wireless resources may lead to resource allocation conflicts, especially when wireless resources are relatively scarce. In this paper, we propose a distributed clustering and resource scheduling scheme based on collaborative perception of wireless resources. The scheme innovatively integrates wireless resource sensing states into the clustering process. Based on the constructed vehicle clustering structure, the cluster heads fuse the distributed perception data of wireless resources from cluster members to generate more accurate collaborative perception results of the environment where the clusters are located, so as to support more reliable resource scheduling and reduce resource allocation conflicts. Simulation results show that the proposed scheme achieves higher topological stability than conventional clustering schemes based solely on mobility, and superior communication performance compared to the autonomous resource allocation mechanism of C-V2X.**

## I. Introduction

Vehicle-to-Everything (V2X) communication is one of the key technologies in the Intelligent Transport System (ITS) [1]. It enhances driving safety, reduces congestion and improves traffic efficiency by sharing driving status information through collaborative communication between vehicles or between vehicles and infrastructure. To meet these diverse communication requirements, Cellular Vehicle-to-Everything (C-V2X) has been introduced and widely adopted as the mainstream technical framework. The first version of the C-V2X standard (i.e., LTE-V2X) introduced modes 3 and 4 in 3GPP Release 14/15 to support direct communication. Mode 3 adopts a resource centralized scheduling scheme, where the base station allocates resources to the vehicles. And mode 4 adopts a distributed resource self-selection scheme, where the vehicles autonomously select resources through a sensing-based semi-persistent scheduling (S-SPS) scheme [2].

Existing research indicates that due to the limitations of the Mode 4 scheme in perception and scheduling, resource allocation collisions may occur in some cases, especially when wireless resources are scarce. References [3] and [4] indicated that since the SPS scheme selects radio resources relying solely on signal strength, it results in inefficient resource utilization and suffers from a lack of adaptability in dynamic or congested environments, particularly under heavy traffic loads or fast-fading channel conditions.

To alleviate the competition over resources, researchers introduced the clustering concept in the vehicle networking [5]. Vehicles are classified into multiple cluster structures based on certain rules. The cluster head usually acts as the local coordinator of its cluster structure, responsible for transmission scheduling, resource allocation, data forwarding, etc. [6]. Resource scheduling based on a clustering structure possesses inherent advantages regarding the reliability of resource allocation, as it enables centralized information aggregation and unified group scheduling of wireless resources [7][8].

Currently, the vast majority of existing studies on cluster-based wireless resource allocation rely on centralized management executed by cellular base stations [9]-[12]. For

This work was supported in part by the Hubei Provincial Science and Technology Program Project under Grant 2025BAA002 and 2024BAA005, in part by Wuhan Institute of Technology under Grand K2023091, and in part by Graduate Innovative Fund of Wuhan Institute of Technology under Grand CX2025309.

instance, in reference [9], a centralized resource allocation scheme based on graph theory is proposed, which is only used for V2V mode 3 communication in overlapping clusters. It adopts a resource allocation strategy in descending order of cluster capacity and a vehicle pre-grouping strategy to avoid temporal domain conflicts within clusters. Sehla et al. [10] proposed a specific inter-cluster resource reuse scheme, which replaces traditional vehicle-to-vehicle reuse with inter-cluster resource reuse and decides whether to reuse the same resource set by calculating the distance between cluster centroids. Reference [12] proposes an improved multi-hop vehicle-mounted communication clustering model based on infrastructure, which comprehensively considers factors such as distance, reliability, and driving speed to optimize network throughput. Although centralized resource scheduling in cluster structures has made significant progress in improving packet delivery rates, such algorithms rely on centralized management of base stations. Furthermore, the scheduling process incurs additional overhead in terms of wireless resources and latency, and scheduling reliability is constrained by imperfect vehicle channel state information, making it difficult to apply these methods to safety-critical service transmissions.

To overcome the limitations of centralized resource scheduling, recent studies have focused on adopting distributed or semi-distributed resource scheduling modes. For example, in references [13] and [14], the resource pool is divided into multiple fixed resource sets and the resource set with the lowest average signal reception intensity is selected based on the wireless resource perception of the cluster head. However, relying solely on the perception results of the cluster head to select the resource set still has certain limitations in vehicle rapid movement scenarios.

In this paper, we propose a distributed clustering and resource selection method based on cooperative resource sensing. By integrating the collaborative perception information of wireless resources into the process of cluster generation and resource allocation, the reliability of distributed resource allocation without base station management under the clustering structure is enhanced. The node with the lowest cluster head factor among neighboring nodes becomes the cluster head and determines whether to include the neighboring nodes in its cluster by calculating the cluster entry factor of each neighboring node. The cluster head selects the resource set with less interference based on the resource coordination perception with the cluster members, and schedules resources for the members within the cluster to reduce resource conflicts.

The remainder of this paper is structured as follows. Section II establishes the comprehensive system model. Section III elaborates on the proposed dynamic vehicle clustering process. Following that, Section IV dilates upon the collaborative-perception-based resource scheduling algorithm. Performance evaluation and quantitative simulation results are presented and analyzed in Section V. Finally, Section VI concludes the paper.

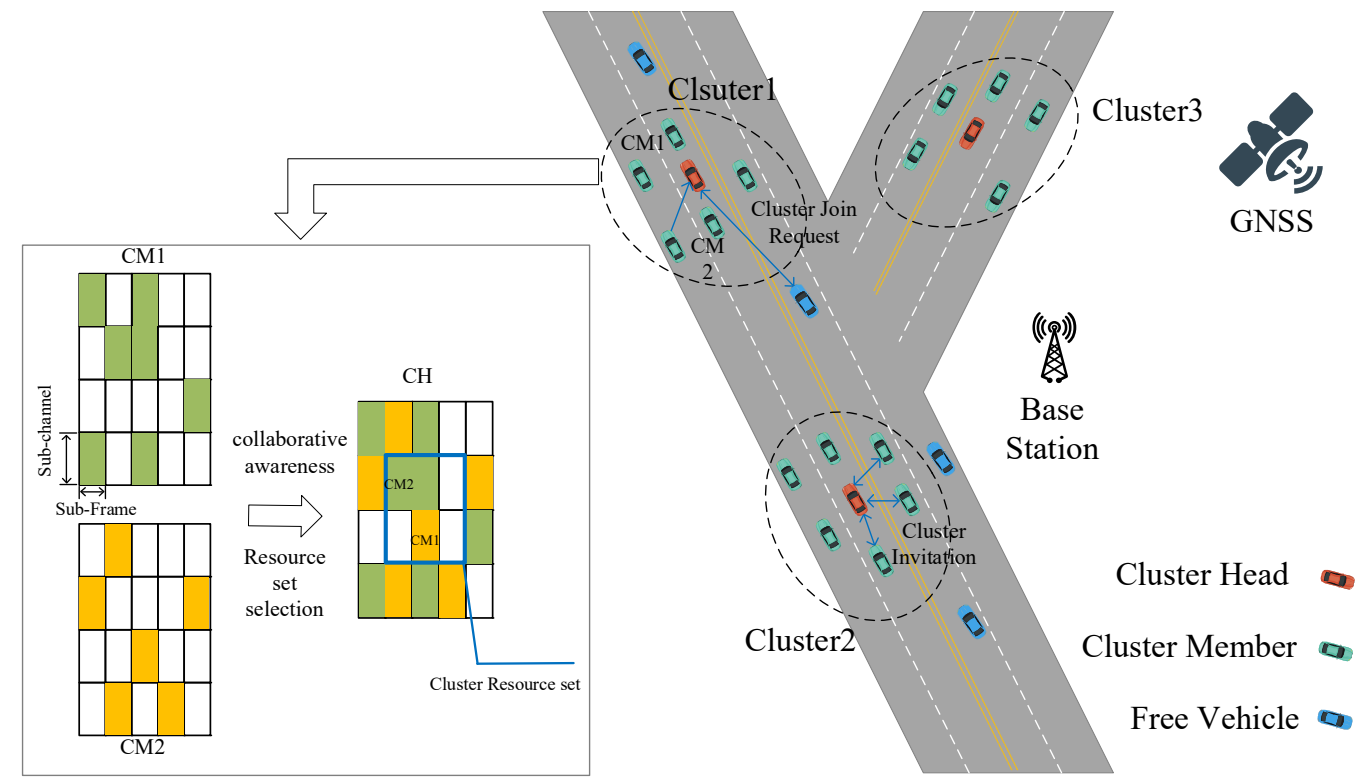

Fig. 1. Cluster Structure

## II. SYSTEM MODEL

The system comprises a set of vehicles $V = \{V_1, V_2, ..., V_n\}$. Each vehicle $V_i \in V$ is equipped with an onboard C-V2X communication unit supporting sidelink communication. The state of vehicle $V_i$ at time t is characterized by its position $P_i(t)$, velocity $V_i(t)$, and a unique identifier (e.g., MAC ID). Through periodically broadcast Beacon messages or Cooperative Awareness Messages (CAMs), each vehicle maintains a dynamic neighbor table $N_i$, which records the status information of all neighboring vehicles.

As shown in Fig. 1, the system defines certain nodes as cluster head (CH), cluster members (CM), and free vehicle (FV). Vehicles with high correlation of mutual mobility and good communication quality are grouped together as a cluster, and each cluster does not overlap with others. Cluster head is designated as the node that achieves the minimum composite score based on mobility metrics and resource utility metrics. Free nodes can select a suitable cluster to apply for membership based on their own movement and communication status, while the corresponding cluster head determines whether to admit the node as a cluster member. Cluster members are exempt from the task of resource selection; the cluster head is responsible for selecting a resource set for the entire cluster and allocating specific resources within that set to all its members.

In C-V2X sidelink communications, single-carrier frequency Division Multiple Access (SC-FDMA) is adopted, and its wireless resource pool is composed of time slots in the time domain and sub-channels in the frequency domain. In the system architecture presented in this paper, the wireless resource pool is modeled as a two-dimensional grid composed of time slots and frequency sub-channels. To facilitate resource scheduling within a clustered structure in a distributed environment, the system employs a Time Division Multiplexing (TDM) framework to partition the C-V2X resource pool into multiple resource subsets. For instance, a 100ms resource pool is divided into five orthogonal resource subsets, each of which consists of continuous time slots and sub-channels, that is, $S = \{S_1, S_2, S_3, S_4, S_5\}$. All nodes within the same cluster will use the same subset of resources.

## III. CLUSTERING SCHEME BASED ON COLLABORATIVE PERCEPTION OF WIRELESS RESOURCES

In this section, an explanation will be provided on how vehicles autonomously form clusters, encompassing the election of cluster head as well as the generation and maintenance of clusters.

### *A. Cluster generation*

When a cluster is formed, the free nodes construct the vehicle movement stability factor *M(i)* based on the speed and distance in the neighbor table, to represent the mobility assessment index of the cluster head. *M(i)* can be calculated as:

$$M(i)=\omega_1 \frac{v_i\text{-}\sum_{j=1}^{N_i} v_j/N_i}{v_{max}\text{-}v_{min}}+\omega_2 \frac{\sum_{j=1}^{N_i} d_{i,j}/N_i}{R_{max}} \tag{1}$$

In the formula, $v_i$ represents the moving speed of node $i$, $d_{i,j}$ is the Euclidean distance between node i and its neighboring node $j$, and $N_i$ is the number of neighboring nodes of node $i$ within the maximum reference transmission radius $R_{max}$. $v_{max}$ and $v_{min}$ are the maximum and minimum speeds of node $i$'s neighboring nodes, respectively, and $\omega_1$ and $\omega_2$ are weight factors. This parameter reflects the consistency level of the moving states of the current node and its neighboring nodes. The smaller *M(i)* is, the higher the synchronization of the movement of this node with its neighboring nodes. Therefore, nodes with a smaller *M(i)* are more likely to become cluster heads, making the constructed cluster have better relative mobility stability.

The utility factor *u(i)* of the resource set is utilized to evaluate the availability and utilization efficiency of the wireless resources around the node. Node *i* calculates the average signal strength of the candidate resource set based on the collaborative perception of the wireless resources of all its neighbor nodes, and jointly assesses the availability and utilization efficiency of the surrounding wireless resources with the number of neighbor nodes. *u(i)* can be calculated as:

$$u(i)=\beta_1 \frac{\sum_{j=1}^{N_i}\left(\sum_{m=1}^{N_T}\sum_{n=1}^{N_F} P(j)_{R_{m,n}}/m\cdot n\right)}{P_{max}\cdot N_i}+\beta_2 \frac{N_{max}\text{-}N_i}{N_{max}} \tag{2}$$

In the formula, $R_{m,n}$ represents the time - frequency position of the resource block. $N_T$ and $N_F$ denote the number of resources in the time domain and frequency domain within the candidate resource set respectively. $P(j)_{R_{m,n}}$ represents the signal reception strength of the resource block $R_{m,n}$ perceived by the neighbor node $j$. $P_{max}$ represents the maximum signal reception strength of neighbor nodes detected by the current node. $N_i$ is the number of neighbor nodes of the current node. $N_{max}$ is the maximum number of neighbor nodes estimated based on the spectrum sensing of neighbor nodes. A smaller *u(i)* indicates that there is a set of wireless resources that are less occupied or reusable near the current location of the vehicle node, meaning that the resource set has high availability. Therefore, a node with a smaller resource utility factor is more qualified to serve as the cluster head.

The cluster head is elected by the node with the smallest cluster head factor $\gamma(i)$ from the neighbor table. The cluster head factor is obtained by adding the mobility factor and the resource set utility factor, which can be derived as:

$$\gamma(i)=M(i)+u(i) \tag{3}$$

The cluster head will calculate the coefficient required for each adjacent vehicle node to join the cluster. This coefficient $\eta_{i,k}$ includes the speed, distance and communication quality of the neighboring vehicle nodes and the cluster head, which can be derived as:

$$\Phi_{i,k}=\varepsilon_1\left|\frac{v_i\text{-}\bar{v}_k}{\bar{v}_k}\right|+\varepsilon_2\frac{1}{\frac{d_{i,k}}{R_k}+1}+\varepsilon_3\left(1\text{-}\left|\frac{P_{i,k}^{rsrp}\text{-}P^{RSRPth}}{P^{RSRPth}}\right|\right) \tag{4}$$

In the formula, $v_i$ represents the speed of the neighboring vehicle node, $\bar{v}_k$ represents the average speed of the cluster, $d_{i,k}$ represents the distance between the neighboring vehicle node and the cluster head, $R_k$ represents the range of the cluster, $P_{i,k}^{rsrp}$ represents the received power of the reference signal between the neighboring vehicle node and the cluster head, and $P^{RSRPth}$ represents the preset threshold of the received power of the reference signal. $\varepsilon_1$, $\varepsilon_2$ and $\varepsilon_3$ are weighting factors. Then, $\Phi_{i,k}$ is compared with the threshold $\delta(i)$ for the cluster, and the neighboring vehicle nodes where $\eta_{i,k}$ is less than $\delta(i)$ are incorporated into the same cluster as itself. The calculation formula for $\delta(i)$ is

$$\delta(i)=\mu\frac{N_i}{N_{max}} \tag{5}$$

Where $N_i$ represents the current number of members in the cluster, and $N_{max}$ denotes the maximum number of cluster members. $\mu$ is weighting factor.

### *B. Cluster maintenance*

This section will describe the behaviors of the cluster head, cluster members, and free nodes after the cluster is generated.

- The behavior of the FV: These FV nodes are the remaining ones after the cluster is generated. When FV receives a message sent by a cluster head, it will calculate the joining-factor of itself and the cluster, and compare it with the joining-threshold sent by the cluster head. If the joining-factor is less than the joining-threshold, it will send a request to join the cluster to the cluster head.
- The behavior of the CH: When the cluster head has not received any messages from the cluster members for a long time, the cluster head will consider that the cluster member has left the current cluster structure. It will remove the cluster member from the cluster member table and notify it in the next Beacon message.
- The behavior of the CM: When a cluster member fails to receive any messages from the cluster head for a long time, the member will consider that it has separated from the current cluster structure and will change its status flag to FV.

To clearly present the execution flow and deployment logic, the proposed decentralized clustering and maintenance mechanism is summarized in Algorithm 1.

**Algorithm 1:** Distributed Cluster Generation and Maintenance Algorithm

**Input:** Neighbor set $N_i$, mobility factor $M_i$, resource utility factor $u_i$.
**Output:** Dynamic cluster states and topology.

**Phase I: Cluster Head (CH) Election**
Each node $i$ initializes $N_i$ and calculates its cluster head factor:$\gamma_i = F(M_i, u_i)$;
**if** $\gamma_i$ is the smallest within its nearby neighborhoods **then**
Node $i$ becomes $CH$;
$CH$ calculates the join threshold $\delta_{th}$;
**for** each neighbor $m \in N_i$ **do**
  $CH$ calculates the factors for joining the cluster $\Phi$;
  **if** $\Phi < \delta_{th}$ **then**
    $CH$ updates the cluster table;
**else**
Node $i$ remains as FV;
**end if**
**Phase II: Periodic Cluster Maintenance**
**Upon** expiration of the maintenance timer **do**
**for** each member $m$ in Cluster **do**
  **if** $CH$ fails to receive the message from $m$ within the time threshold **then**
    Evict $m$ from cluster;
  **end if**
**end for**
Update dynamic join threshold $\delta_{th}$;
**end Upon**

## IV. ALLOCATION BASED ON WIRELESS RESOURCE COLLABORATIVE PERCEPTION

The cluster head evaluates the availability of the cluster resource set based on the collaborative wireless resource sensing results from cluster members and assigns specific wireless resources to each cluster member, achieving semi-centralized scheduling of wireless resources based on the clustering structure. Compared with the centralized resource scheduling based on base stations, it has better flexibility. Compared with the perception results of cluster heads themselves, the collaborative perception results of wireless resources based on cluster members can more accurately reflect the wireless environment monitoring around the cluster.

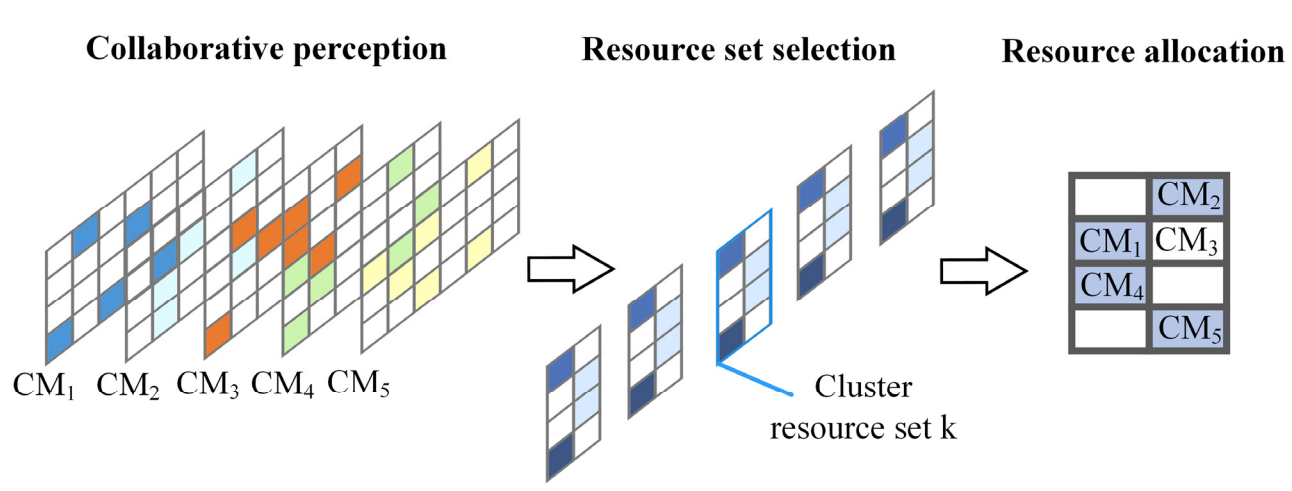

Fig. 2. scheduling of resources

### A. Resource collaborative perception

Each vehicle node acts as a distributed sensing node, continuously monitoring the wireless channels within the sensing window and measuring the received power of the reference signals. Once it becomes a cluster member, this node will send the local sensing results to the cluster head through Beacon messages. After receiving the sensing information from each cluster member, the cluster head will construct a collaborative resource sensing map and use this map to further select the resource set and allocate resources to each cluster member.

### B. Cluster Head selection resource set

The cluster head will select the resource set based on the comprehensive resource collaborative perception map. We set a resource selection interval of once per second, and select the resource set with the minimum average RSRP value based on the collaborative perception results as the resource set to be used for the subsequent operations of the cluster. The formula for calculating the average RSRP of the resource set is as follows:

$$\overline{P_{R_{m,n}}} = \sum_{j=1}^{N_K} P(j)_{R_{m,n}} \Big/ N_K \tag{6}$$

In the formula, $N_K$ represents the number of cluster members in cluster $K$, and $P(j)_{R_{m,n}}$ indicates the signal reception strength of resource block $R_{m,n}$ perceived by cluster member $j$.

### C. Cluster Head allocates resources to the cluster members

After the cluster head selects the resource set, it will allocate resources to each cluster member. Here, we divide resource allocation into initial allocation and subsequent allocation. The initial allocation occurs when the cluster structure is first established, at which point the cluster head will allocate resources to all cluster members (including itself). Subsequent allocation refers to other resource allocations apart from the initial allocation. A resource retention probability $P_{keep}$ similar to that in C-V2X Mode 4 is set to control the reallocation of cluster resources. When the random probability is less than 1- $P_{keep}$, the cluster head will re-select the resource set and re-allocate resources to the cluster members. Additionally, in the subsequent allocation, we adopt a local allocation strategy and only perform resource redistribution for cluster members with a high transmission failure rate. The transmission failure rate can be expressed as:

$$TR_i = \frac{N_i^{Failure}}{N_i^{Total}} \tag{7}$$

Where $N_i^{Failure}$ represents the number of messages sent by cluster member $i$ to the cluster head but which the cluster head fails to parse, while $N_i^{Total}$ represents the total number of messages sent by cluster member $i$ to the cluster head.

When allocating specific resource blocks to cluster members, the cluster head will comprehensively evaluate the collaborative perception results of this member and other cluster members regarding the resource block. Only when the cumulative perception intensity of a certain resource block is lower than that of other resource blocks within the candidate resource set, will the cluster head allocate this resource block to the member.

To clearly present the operational flow of the cluster-based resource management, the detailed execution logic of the proposed resource allocation is summarized in Algorithm 2.

**Algorithm 2:** Cluster-based Distributed Resource Allocation Algorithm

**Input:** Cluster member set $M_k$, Candidate resource pool $S = \{R_{m,n}\}$, $P_{keep}$.
**Output:** allocated resource block $R_i$ for member $i$.

1. **Phase I: Cluster Resource Set Selection**
2. Partition candidate pool $S$ into distinct resource sets$\{S_1, S_2, S_3, S_4, S_5\}$.
3. Calculate the average RSRP for each $S_i$ based on collaborative sensing results.
4. Select the resource set $S_{selected}$ with the minimum average RSRP.
5. **Phase II: Resource Block Allocation**
6. **if** Initial Allocation (Cluster formation) **then**
7.   **for** each cluster member $i \epsilon M_k$ **do**
8.     Calculate the cumulative perception intensity of each $R_{m,n} \in S_{selected}$;
9.     Allocate the $R_{m,n}$ with the minimum cumulative perception intensity to $i$;
10. **end for**
11. **else**(Subsequent Allocation)
12.   Generate a random uniform probability $p \epsilon [0,1]$;
13.   **if** $p < 1 - P_{keep}$ **then**
14.     Re-execute Phase I to update;
15.     **for** each cluster member $i \epsilon M_k$ **do**
16.       **if** transmission failure ratio $\rho_i$ is high **then**
17.         Execute lines 8 to 9;
18.       **else**
19.         Retain the original resource;
20.       **end if**
21.     **end for**
22.   **end if**
23. **end if**

## V. SIMULATION RESULT

The proposed scheme is thoroughly compared with the benchmark 3GPP Mode 4 S-SPS, Resource-Blind Clustering, and Mobility-based Scheme for Dynamic Clustering (MSDC) [15] algorithms to demonstrate its superiority in both packet collision ratio and cluster topological stability. Specifically, MSDC is a mobility-aware scheme that forms clusters based on vehicle movement directions, relative positions, and estimated link lifetimes, while employing a safe distance threshold to control cluster sizes. Since the original MSDC solution does not involve a resource scheduling scheme, we combine its clustering scheme with our resource scheduling scheme. Resource-Blind Clustering serves as a crucial ablation baseline derived from our proposed scheme. It utilizes the same clustering framework but deliberately disengages all resource-related metrics during the cluster formation phase, while retaining the identical resource scheduling strategy. All the proposed clustering algorithms and resource scheduling algorithms have been implemented on the OMNeT++ platform, and the traffic simulation scenarios were generated by the Simulation of Urban MObility (SUMO).

In the test scenario, the road topology consists of a three-lane two-way expressway with a length of 3 kilometers. The vehicles are generated according to the pre-set traffic flow, with the maximum speed being 38.89 meters per second. Traffic flow densities of 0.06 veh/m, 0.12 veh/m, 0.18 veh/m were considered. The transmission period of each Beacon message is 100ms, its size is 190Bytes and it occupies one sub-channel. The other simulation parameters are provided in Table 1.

TABLE I. SIMULATION PARAMETERS

| Parameters | Value |
|---|---|
| Vehicle density | 0.06,0.12,0.18 |
| Road length | 3 km |
| Number of lanes | 3 lanes in each direction |
| Beacon message size | 190 Byte |
| Transmission interval | 100 ms for Beacon |
| Carrier frequency | 5.89 GHz |
| Channel bandwidth | 10 MHz |
| Transmission power | 23 mW |
| Reference transmission radius | 500 m |
| Number of subchannels in each subframe | 4 |
| Size of subchannel | 12 RBs |

### A. *Performance metrics*

The metrics used for evaluating clustering stability and resource scheduling performance are as follows

- Average CH duration: This indicator represents the lifecycle of a cluster head, which is the time interval between the moment when a vehicle enters the Cluster Head (CH) state and the moment when it transitions to another state. Generally, a longer CH duration indicates a more stable cluster.
- Average CM duration: This indicator defines the average time interval between a node joining an existing cluster as a CM and its subsequent exit from the cluster or transformation into a CH.
- Packet Collision Ratio (PCR): This indicator is described as the ratio of the number of data packets lost due to transmission conflicts to the total number of packets sent. The lower this indicator is, the better the communication quality is.

### B. *Results and analysis*

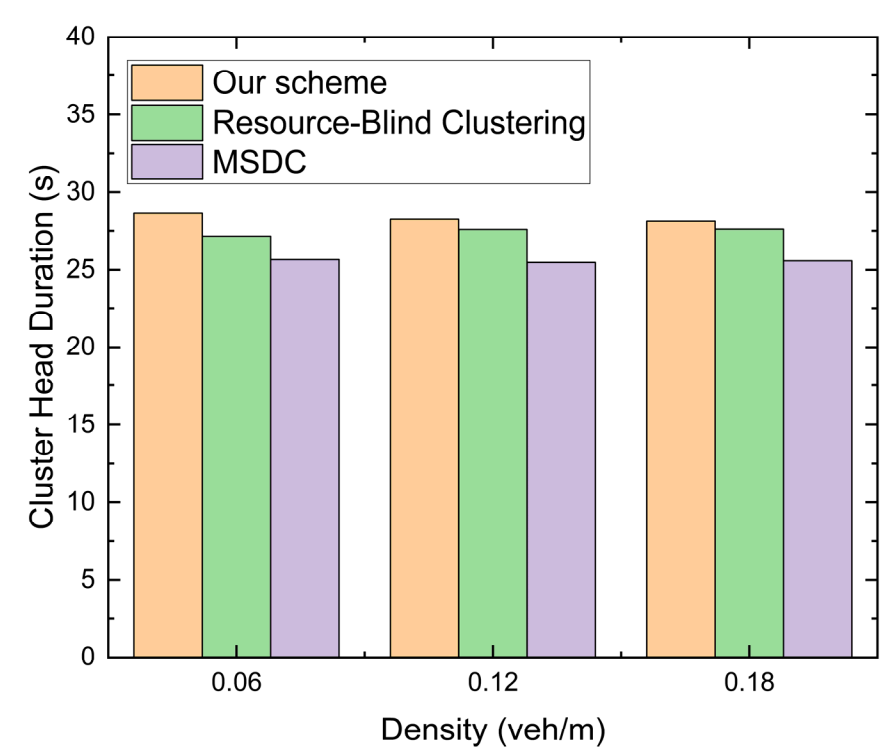

Fig. 3. Comparison of the holding time of cluster heads at various densities

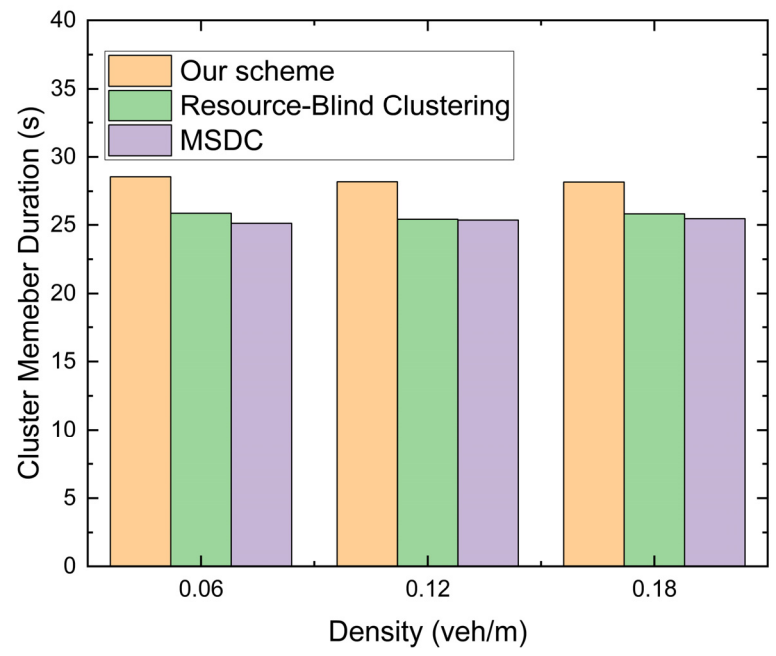

Fig. 4. Comparison of the holding time of cluster members at various densities

Fig. 3 and Fig. 4, the comparisons of the cluster head maintenance time and cluster member maintenance time in the proposed scheme and the MSDC scheme and Resource-Blind Clustering scheme are respectively presented. As shown in the figure, the proposed scheme outperforms the MSDC and Resource-Blind Clustering schemes in both cluster head maintenance time and cluster member maintenance time. This indicates that this scheme exhibits higher stability in terms of cluster structure stability. During the clustering phase, we not only take mobility and geographical location into account but also incorporate resource indicators. Taking the availability of resource sets into account during cluster head selection enables the cluster head to retain available resources for a certain period, thereby allowing it to maintain its role for a longer time. When selecting cluster members, an entry threshold is maintained and the communication quality between cluster members and cluster heads is taken into account, enabling cluster members to establish more stable connections with clusters.

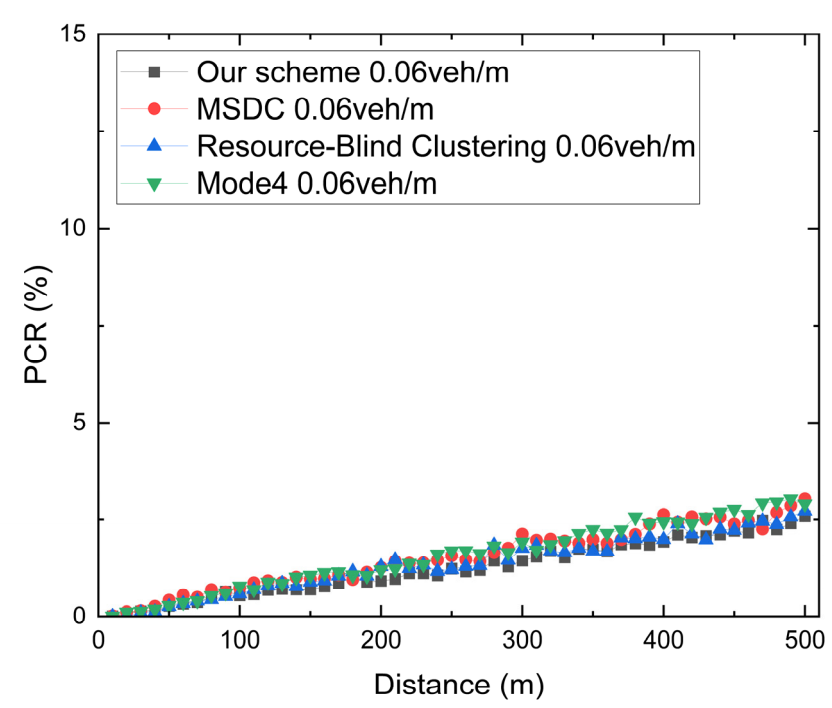

Fig. 5. Comparison of PCR at a density of 0.06 veh/m

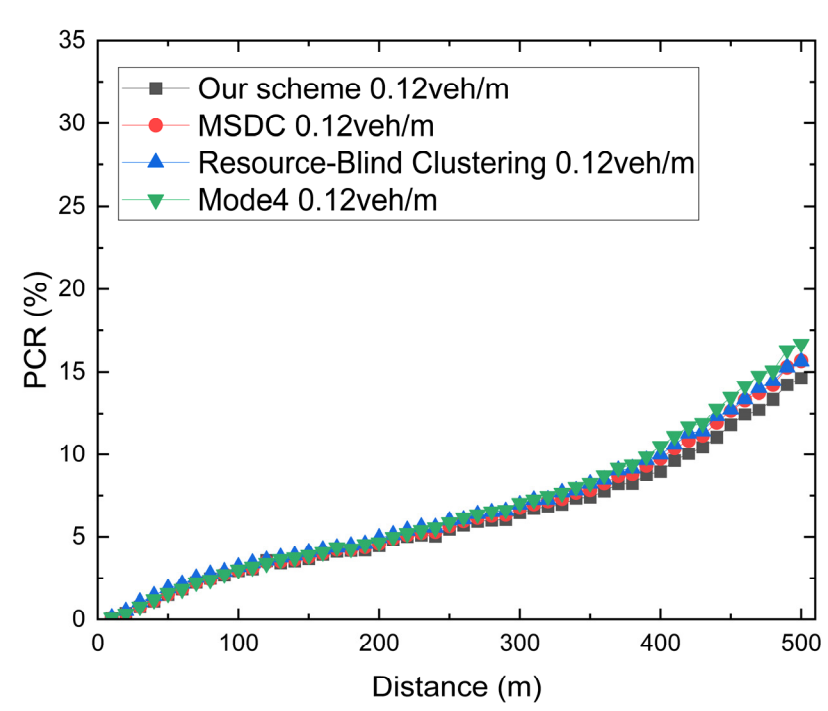

Fig. 6. Comparison of PCR at a density of 0.12 veh/m

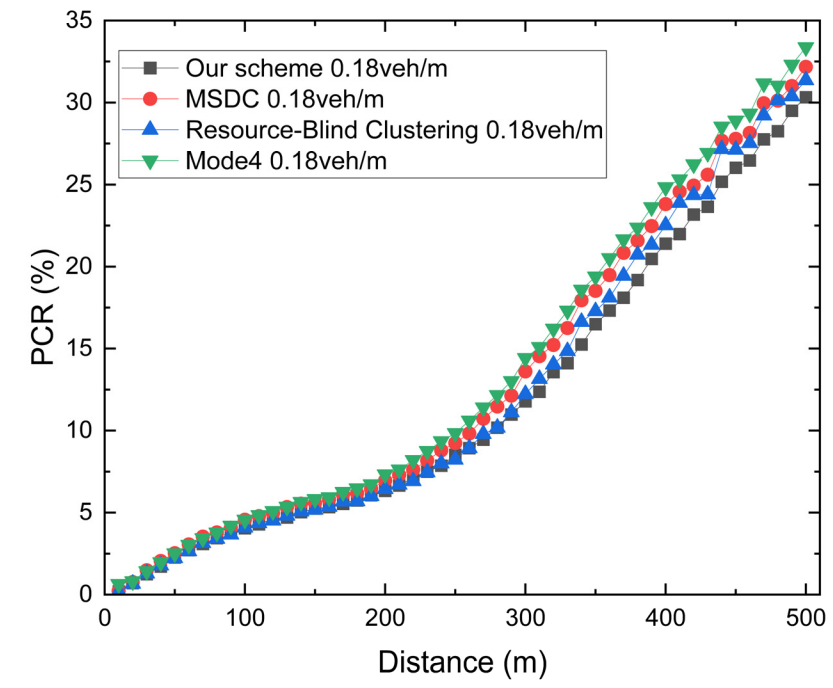

Fig. 7. Comparison of PCR at a density of 0.18 veh/m

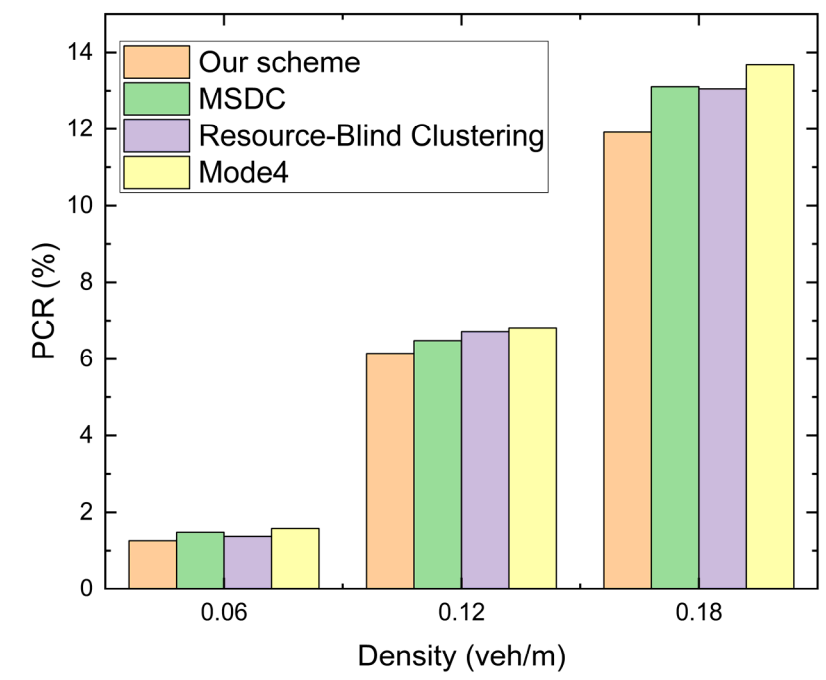

Fig. 8. Comparison of the mean values of PCR at different densities

Fig. 5, Fig. 6 and Fig. 7 compared the changes in packet collision ratio with distance under the densities of 0.06, 0.12, and 0.18 veh/m for the proposed scheme, as well as for the MSDC, Resource-Blind Clustering and C-V2X mode 4. In Fig. 8, the average PCR values of the three schemes under different densities are compared. As can be seen from the figure, the PCR values of all four schemes gradually increase with the increase in distance. However, the PCR value of our proposed scheme is lower than those of the other three schemes in most cases, particularly maintaining the lowest average PCR across all tested vehicle densities. At a density of 0.06 veh/m, since the resources are relatively abundant, the PCR values of all three schemes are relatively low. The improvement effect of our proposed scheme is not significant. The average PCR value of our scheme is only 0.273%, 0.11% and 0.313% lower than the MSDC, Resource-Blind Clustering and C-V2X mode 4, respectively. At a density of 0.12 veh/m and 0.18 veh/m, the channel resources start to become increasingly strained, and the superiority of our proposed solution start to manifest. At a distance of 500 meters under 0.18 veh/m, the PCR of our proposed solution compared to MSDC, Resource-Blind Clustering and C-V2X reduced by approximately 1.85%, 1.64% and 3.05% respectively. The average value of PCR decreased by 1.19%, 1.14% and 1.76%. The reasons for the above results can be summarized as follows: 1) Compared with MSDC, our proposed scheme has better stability in the cluster structure and a more reasonable selection of cluster heads and cluster members. Nodes with better quality available resource sets become cluster heads, enabling cluster heads to better complete the subsequent rational allocation of resources. 2) Our proposed resource scheduling scheme based on collaborative perception, since the cluster head integrates the perception views of each cluster member, it has a more comprehensive resource perception view and can better exclude resource blocks with a higher collision probability. At the same time, the cluster head uniformly allocates resources for each cluster member, achieving orthogonal resources within the cluster and further reducing the probability of resource conflicts.

## VI. CONCLUSION

This paper first proposes a distributed cluster scheme of C-V2X, integrating the availability of wireless resources into the clustering process. It constructs a joint objective function for cluster head election and node joining, to improve the stability of the cluster and provide a good resource foundation for the reliability of subsequent resource allocation. Next, a semi-centralized wireless resource scheduling algorithm based on a cluster-based structure is proposed for scenarios without base station management. It utilizes the collaborative perception of wireless resources among cluster members to enhance the overall perception accuracy of the cluster , thereby improving the reliability of resource allocation within the cluster. Simulation results demonstrate that the proposed scheme significantly improves cluster stability and communication reliability, offering a promising approach for C-V2X communication.